\documentclass{WileyMSP-template}
\usepackage{graphicx}
\usepackage{dcolumn}
\usepackage{multicol}
\usepackage{braket}
\usepackage{float}
\usepackage{bm}
\usepackage{amsmath}
\usepackage{cite}
\usepackage{xcolor}
\usepackage{amssymb}
\usepackage{gensymb}
\usepackage{dirtytalk}
\usepackage [english]{babel}
\usepackage [autostyle, english = american]{csquotes}
\usepackage{textcomp}
\MakeOuterQuote{"}

\begin{document}

\rhead{\includegraphics[width=2.5cm]}

\title{Device-independent, megabit-rate  quantum random number generator with beam-splitter-free architecture and live Bell test certification}

\maketitle

\vspace{12pt}
\author{Ayan Kumar Nai*},
\author{Vimlesh Kumar},
\author{M. Ebrahim-Zadeh},
\author{G. K. Samanta}





\vspace{12pt}
\begin{affiliations}
Ayan Kumar Nai\\
Photonic Sciences Lab., Physical Research Laboratory, Ahmedabad 380009, Gujarat, India\\
Indian Institute of Technology Gandhinagar, Ahmedabad 382424, Gujarat, India\\
Email Address: ayankrnai@gmail.com\\
\vspace{12pt}

Vimlesh Kumar, Prof. G. K. Samanta\\
Photonic Sciences Lab., Physical Research Laboratory, Ahmedabad 380009, Gujarat, India\\
\vspace{12pt}
Prof. M. Ebrahim-Zadeh\\
ICFO-Institut de Ciencies Fotoniques, 08860 Castelldefels (Barcelona), Spain\\
Institucio Catalana de Recercai Estudis Avancats (ICREA), Passeig Lluis Companys 23, Barcelona 08010, Spain\\
\vspace{12pt}

\end{affiliations}



\justifying
\begin{abstract}
Device-independent quantum random number generators (DI-QRNGs) are crucial for information processing, ensuring certified quantumness and genuine randomness. However, existing implementations often face low bit rates due to quantumness testing challenges.  Here, we present a high-bit-rate DI-QRNG with live quantumness certification through the Bell test. Using spontaneous parametric down-conversion in a polarization Sagnac interferometer, we generate entangled pair-photons at diametrically opposite points on an annular ring with strong spatial and temporal correlations. Dividing the ring into six diametrically opposite sections, we create three entangled photon sources that exhibit bias-free quantum mechanical randomness from a single resource.  By utilizing the coincidence counts of pair-photons from two sources, we generate raw bits, while the third source simultaneously measures the Bell's parameter without any loss of QRNG bits. We have generated 90 million raw bits in 46.4 seconds with the Bell parameter (S $>$ 2), with a minimum entropy extraction ratio exceeding 97$\%$. Post-processed using a Toeplitz matrix, the DI-QRNG achieves a bit rate of 1.8 Mbps, passing all NIST 800-22 and TestU01 tests. In the absence of Bell's parameter for a non-maximally entangled state, $g^{(2)}(0)$ can be the metric for quantumness measure. Scalable and beam-splitter-free, this megabit-rate DI-QRNG is ideal for practical applications.

\end{abstract}

\begin{multicols}{2}
\section{Introduction}
\justifying
\noindent Random numbers play a fundamental role in a wide range of applications, including cryptographic protocols, secure communication, and scientific simulations \cite{Miguell2017, James2020}. The random numbers generated through the deterministic classical computer algorithms, known as the pseudo-random number generators (PRNGs), are vulnerable to predictability and external tampering \cite{Yev, Bras}. On the other hand, the random numbers generated by exploiting the inherent unpredictability of quantum stochastic process, known as the quantum random number generators (QRNGs), guarantee the generation of truly random numbers \cite{Ma, Stip}. Typically, the quantum stochastic process includes the measurement of quantum-phase fluctuation, single photon detection, and vacuum fluctuations \cite{Xu:2012, Dynes:2008, Gabriel:2010}. Conventionally, the QRNGs have two parts: source (generating quantum states that are inherently random) and measurement (demolition of quantum state using a practical device such as a beam-splitter and subsequent detection using a detector to generate raw bits). However, in the absence of perfect devices for generation and measurement, practical QRNGs can suffer from security loopholes, leading to the development of device-independent QRNGs (DI-QRNGs) independent (regardless) of the physical implementation and characterization of the instruments used \cite{Pironio:2010, Acin:2016, Liu:2018}.\\
\indent Typically, the DI-QRNGs have the inherent certification of randomness based solely on the violation of Clauser-Horne-Shimony-Holt (CHSH) form of Bell's inequality \cite{CHSH:69} without assuming trust in the internal workings of the quantum random number generation scheme \cite{Leone:2022, Xu:16}. Despite having the highest security in the random numbers, the DI-QRNGs have a relatively lower bit rate \cite{Christensen:2013, Leone:2022} due to the limitations on the brightness of the entangled sources developed using the current technologies. As a result, the practical implementations of DI-QRNGs in parallel with the QRNG scheme have not been explored to date. Alternatively, efforts have been made to enhance the bit rate of the QRNG through exploration of different protocols such as source-independent QRNG\cite{Cao:2016, Marangon:2017, Zheng:2020, Ayan:24}, measurement device-independent QRNG\cite{Yu:2021, Pirandola:2015}, and semi-device-independent QRNG \cite{Li:2024, Cheng:2024} with different assumptions about the source or the measurement at the cost of ultimate security of the randomness. Additionally, the device-independent QRNGs are not loss-tolerant, demanding very severe requirements on experimental devices. On the other hand, the majority of the QRNGs often rely on physical devices such as 50:50 beam-splitters or polarization beam-splitters, despite the technological challenge of making a perfect lossless, unbiased beam-splitter. \\
\indent Therefore, it is imperative to devise new protocols for the development of DI-QRNG with a high bit rate while ensuring randomness by the intrinsically probabilistic nature of quantum physics. Recent advances in quantum optics and engineering have opened pathways to simplify DI-QRNG architectures. Beam-splitter-free designs are particularly promising, offering streamlined setups that reduce complexity and enhance robustness \cite{Chen, Hayl, Ayan:24}. Additionally, incorporating real-time randomness certification, such as live Bell test verification, strengthens the trustworthiness of the generated numbers by providing immediate validation of the CHSH inequality \cite{CHSH:69} for the source.\\
\indent In this paper, we introduce a novel beam-splitter-free, high-bit-rate DI-QRNG with live Bell test certification. Using the non-collinear, degenerate, spontaneous parametric downconversion (SPDC) process in a 20-mm-long periodically-poled potassium titanyl phosphate (PPKTP) crystal in a polarization Sagnac interferometer, we have developed three high-brightness entangled photon sources \cite{Jabir2017,Singh:2023a} from the single resources (optical components, nonlinear crystal, and laser source) by dividing the SPDC ring into six diametrically opposite sections. Since the generation of the pair-photons over the diametrically opposite points of the SPDC ring is purely random and governed by the quantum mechanical process, we have generated random bits (raw) with min-entropy exceeding 97$\%$ by measuring the coincidence of the quantum correlated pair-photons of the two entangled photon sources without any projection and third source for live measurement of CHSH Bell's parameter ensuring real-time trustworthiness of the source, thus confirming the realization of DI-QRNG. The entropy ensures high efficiency in extracting true random bits. Using Toeplitz matrix-based post-processing, our DI-QRNG system results in a bit rate of 1.8 Mbps, passing all NIST 800-22 and TestU01 statistical test suites. Such a generic experimental scheme can be scalable to provide a higher bit rate through the generation of a multi-bit DI-QRNG system with live certification by dividing the SPDC ring to form more entangled photon sources from the single resource.

\section{Experiment}
\begin{figure*}[ht]
    \centering
    \includegraphics[width=\linewidth]{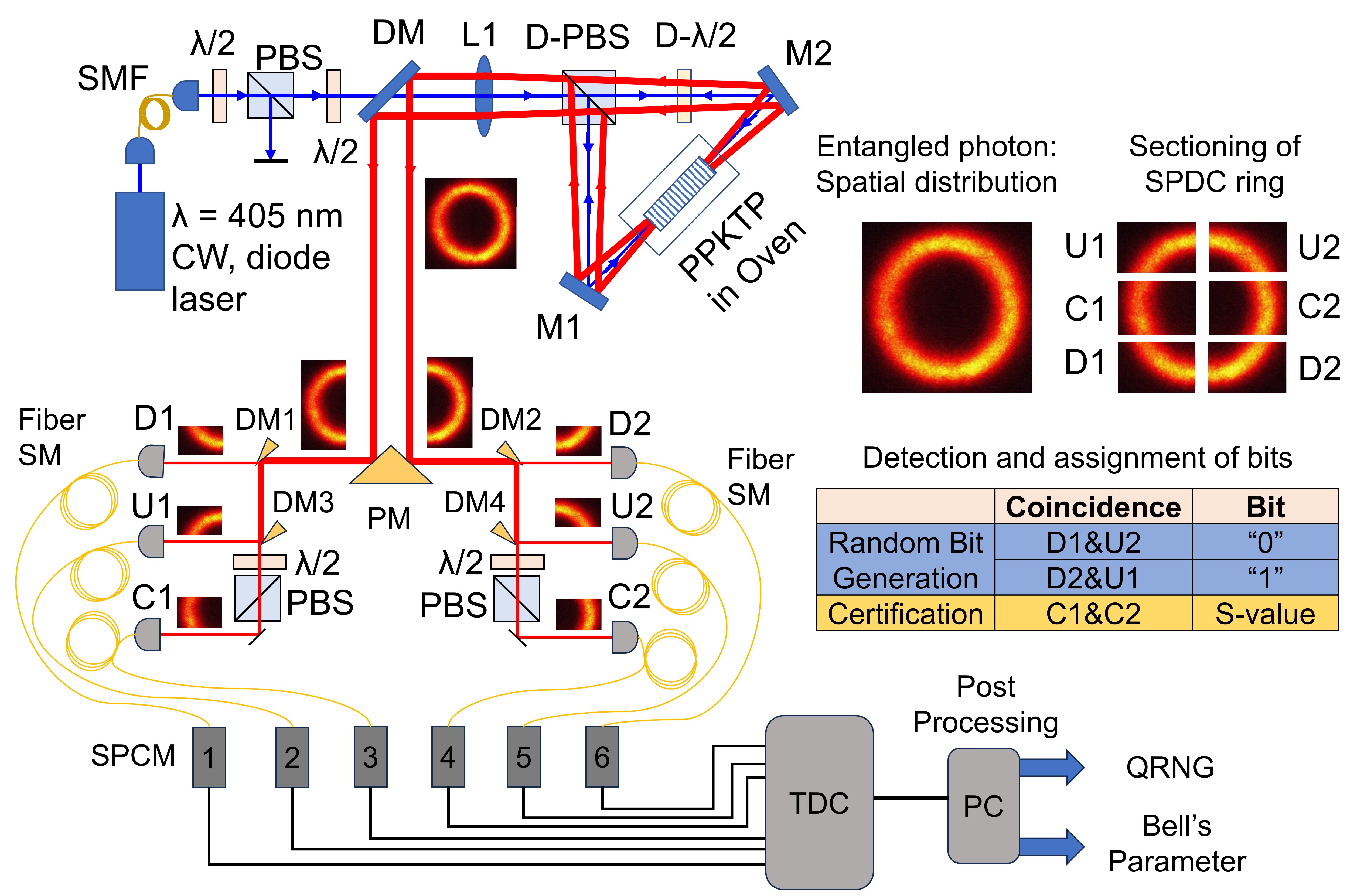}
    \caption{\textbf{The schematic representation of the experimental setup of the QRNG source.} Laser: 405 nm, cw diode laser, $\lambda$/2: Half-wave-plate, L1: plano-convex lenses of focal length 150 mm, $D-\lambda$/2: Dual-half-wave-plate, PPKTP: 20-mm-long periodically-poled potassium titanyl phosphate crystal of single grating in the oven, IF: Interference filter, PM: Prism-shaped gold-coated mirror, DM1-4: D-shaped mirrors, U1-2, D1-2, and S1-2: Upper, lower and side sections of the SPDC ring, SM: Single-mode fiber, SPCM1-6: single photon counting modules, TDC: Time-to-digital converter, PC: Computer. Inset: Spatial distribution of entangled photon source, the architecture of raw bit assignment, and quantumness certification.}
    \label{Figure 1}
\end{figure*}
The schematic of the experimental setup is illustrated in Fig. \ref{Figure 1}. A continuous-wave, single-frequency (linewidth $\sim$20 MHz), single-mode fiber (SMF)-coupled laser diode providing an output power of 21 mW at a central wavelength of 405 nm is used as the pump source. Operating the laser at its highest power to ensure optimum performance, the power attenuator, comprised of a half-wave-plate ($\lambda/2$) and a polarizing beam-splitter (PBS) cube, is used to control the input pump power to the experiment. A second $\lambda/2$ plate with optic axis at an angle $\theta$ with respect to the vertical axis is used to transform the horizontally (H) polarized pump beam in the superposition of horizontal and vertical (V) polarization states in the form of ($\alpha'$ H - $\beta '$ V), where, $\alpha'$ = sin(2$\theta)$, $\beta'$ = cos(2$\theta)$, and $|\alpha'|^2 + |\beta'|^2 = 1 $. As a result, adjusting $\theta$, we can control the laser power in both clockwise (CW) and counterclockwise (CCW) beams of the polarization Sagnac interferometer, which consists of a dual-wavelength PBS (D-PBS, for 405 nm and 810 nm) and two plane mirrors, M1 and M2. The lens, L1, with a focal length of $f$ = 150 mm, focuses the pump beam to a spot size of $\sim$40 $\mu$m at the center of the nonlinear crystal placed between mirrors M1 and M2. The dual-wavelength half-wave-plate (D-$\lambda/2$), with its optic axis rotated by 45$\degree$ relative to the vertical, is placed inside the Sagnac interferometer to transform horizontal polarization to vertical and vice versa for both 405 nm and 810 nm wavelengths. A single-grating ($\Lambda$ = 3.425 $\mu$m) periodically-poled potassium titanyl phosphate (PPKTP) crystal with a length of 20 mm and a 1$\times$2 mm$^2$ aperture serves as the nonlinear medium, producing degenerate SPDC photons at 810 nm through non-collinear, type-0 ($e$ $\rightarrow$ $e$ + $e$) phase-matching. The crystal is housed in an oven with a temperature range up to 200$\degree$C and a stability of $\pm$0.1$\degree$C.\\
\indent The CW and CCW pump beams, in the presence of the D-$\lambda/2$ plate, generate pair-photons distributed in an annular ring with orthogonal polarization states. When superimposed on the D-PBS, the SPDC photons exhibit an annular ring distribution, with pair-photons located at diametrically opposite points on the ring and described by the state, $\ket{\phi^-}$ = $(\alpha \ket{HH} - \beta \ket{VV})$, where, $\alpha$ = sin(2$\theta)$, $\beta$ = cos(2$\theta)$, and $|\alpha|^2 + |\beta|^2 = 1 $. Here, $\alpha$ and $\beta$ are function of $\alpha'$ and $\beta'$. The optic axis of $\lambda/2$ plate at an angle $\theta$ = $22.5\degree$ with respect to horizontal results in SPDC photons with state, $\ket{\phi^-}$ = $(\ket{HH} - \ket{VV})/\sqrt{2}$ (a Bell state). The SPDC photons are collimated by lens L1 and separated from the pump beam by a dichroic mirror (DM) positioned between lens L1 and the $\lambda/2$ plate. The collimated annular SPDC ring, as observed by the EMCCD camera (Andor, iXon Ultra 897) with an interference filter of bandwidth $\sim$3.2 nm centered at 810 nm, is shown by the image in the inset of Fig. \ref{Figure 1}. \\
\indent We divide the SPDC into two halves by the gold-coated prism mirror (PM), and subsequently, each half is divided into three subsections by D-shaped mirrors, DM1-4, resulting in three pairs of diametrically opposite sections (U1, D2), (U2, D1) and (C1, C2) having pair-photons. While the photons from each section are collected using the combination of an interference filter (bandwidth of 3 nm) mounted fiber coupler, single-mode fiber (SM), and single photon counting modules, SPCM1-6, (SPCM-AQRH-14-FC), the photons of sections C1 and C2, pass through the polarization analyzer consisting of a $\lambda/2$ plane and PBS before collection. The time-to-digital converter (TDC) (MultiHarp 150, PicoQuant) is connected to the SPCMs, and the computer is used for photon counting, time stamping, recording, and experimental data analysis. The three pairs of diametrically opposite sections of the annular ring form the three sets of single/entangled photon sources. For raw bit generation, as shown by the table in Fig. \ref{Figure 1}, we assign bits "0" and "1" for the coincidence event between the photons of sections (D1 $\&$ U2) and (D2 $\&$ U1), respectively. As the appearance of pair-photons in sections (U1, D2) and (U2, D1) are quantum mechanically random, this method produces a sequence of random bits of 0 and 1. The photons of sections (C1 and C2) are used for the entanglement study of the pair-photons through the polarization projective measurement and Bell's parameter calculation. The coincidence window throughout the manuscript, unless otherwise mentioned, is 1 ns.




\section{Results and discussions}
\begin{figure}[H]
    \centering
    \includegraphics[width=\linewidth]{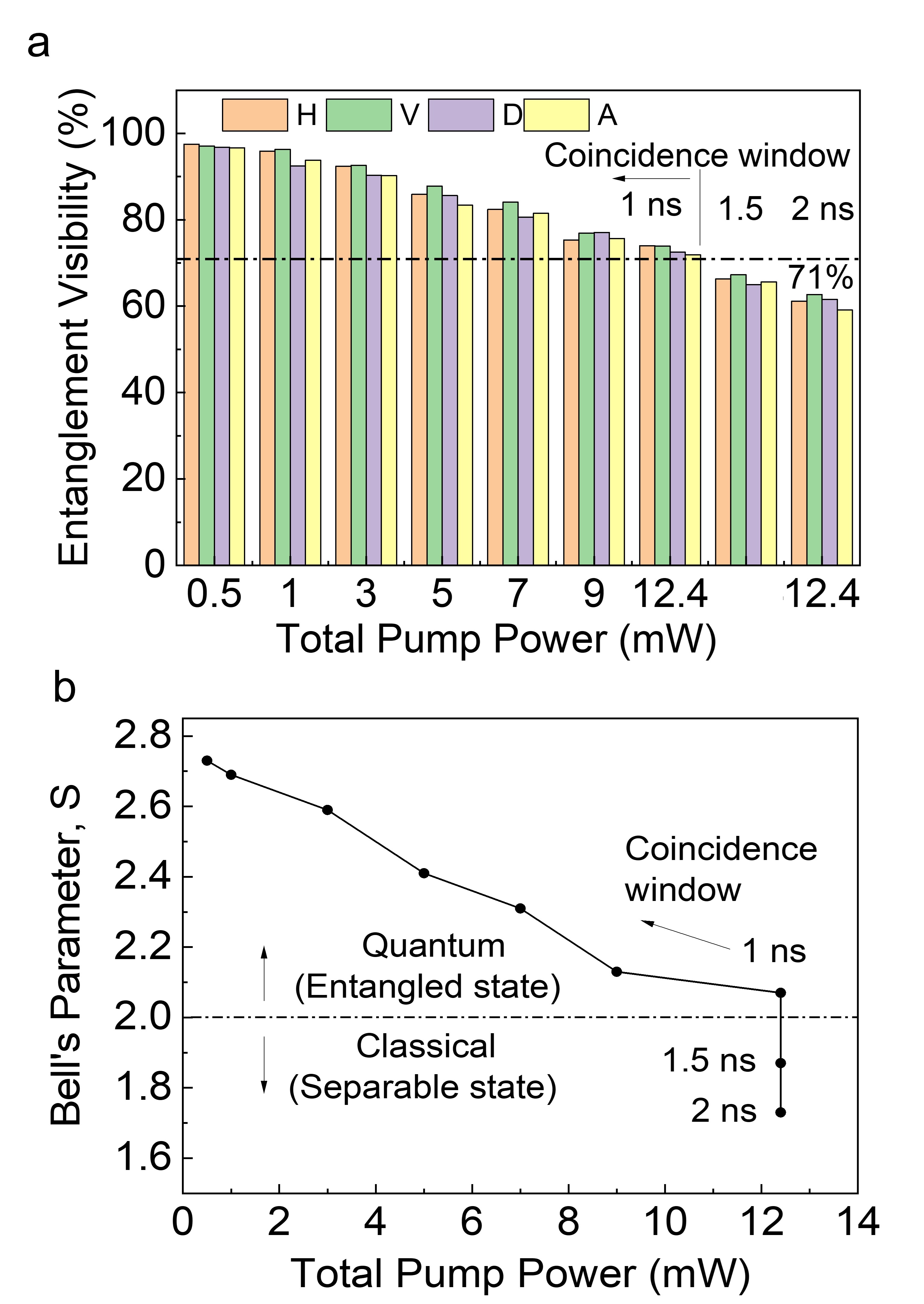}
    \caption{\textbf{Characterization of the entangled photon sources.} Variation of (a) Entanglement visibility and (b) Bell parameter (S) of the entangled photon source as a function of pump power and with coincidence window. }
    \label{Figure 2}
\end{figure}

We first characterized the entangled photon source by measuring the interference visibility in different bases and the value of Bell parameters as a function of pump power and coincidence window. Using the pair-photons from the sections (C1, C2), we have set the polarization state of one of the photons (say from C2) at horizontal (H), vertical (V), diagonal (D), and anti-diagonal (A) using the $\lambda/2$ plane and measured the interference visibility, V = (max. - min.)/(max. + min.), by recording the coincidence counts while continuously rotating the polarization state of the other photon (from C1) using $\lambda/2$. The visibility for different bases as a function of pump power is shown in Fig. \ref{Figure 2}.
\indent As evident from Fig. \ref{Figure 2}(a), the entanglement visibility in all bases (H, V, D, and A) are almost comparable for each pump power, but with the increase of pump power from 0.5 mW to 12.4 mW and coincidence window of 1 ns, the visibility decreases from 97$\%$ to 73$\%$, respectively. However, the interference visibility of $>$71$\%$ for all power levels confirms the entanglement \cite{Jabir2017}. Moreover, keeping the pump power constant at 12.4 mW, we measured the interference visibility further decreasing from 73$\%$ to 61$\%$, below the 71$\%$ limit to ensure entanglement, with the increase of coincidence window from 1 ns to 2 ns with a step of 0.5 ns. The measurement of Bell's parameter, $S$, as shown in Fig. \ref{Figure 2}(b), confirms the decrease of S-value from 2.73 to 2.07 with the pump power increase from 0.5 mW to 12.4 mW. Although S$\geq$2 confirms the entanglement, the increase of the coincidence window from 1 ns to 2 ns drastically reduces the S-value from 2.07 to 1.73 due to the generation of multi-photons at high pump power and detection of such photons in the presence of longer coincidence window ($>$1 ns). Such observation confirms the control transition of the output of the pair-photon source from the quantum state to the classical two-photon state. As expected, we observe the similar performance of the pair-photons collected from the sections (U1, D2) and (U2, D1), confirming the realization of three entangled photon sources from a single set of resources (laser and crystal). In fact, by dividing the annular ring of the photons into more pairs, one can develop a larger number of entangled photon sources; however, this comes at the cost of reduced brightness.\\
%

\begin{figure}[H]
    \centering
    \includegraphics[width=\linewidth]{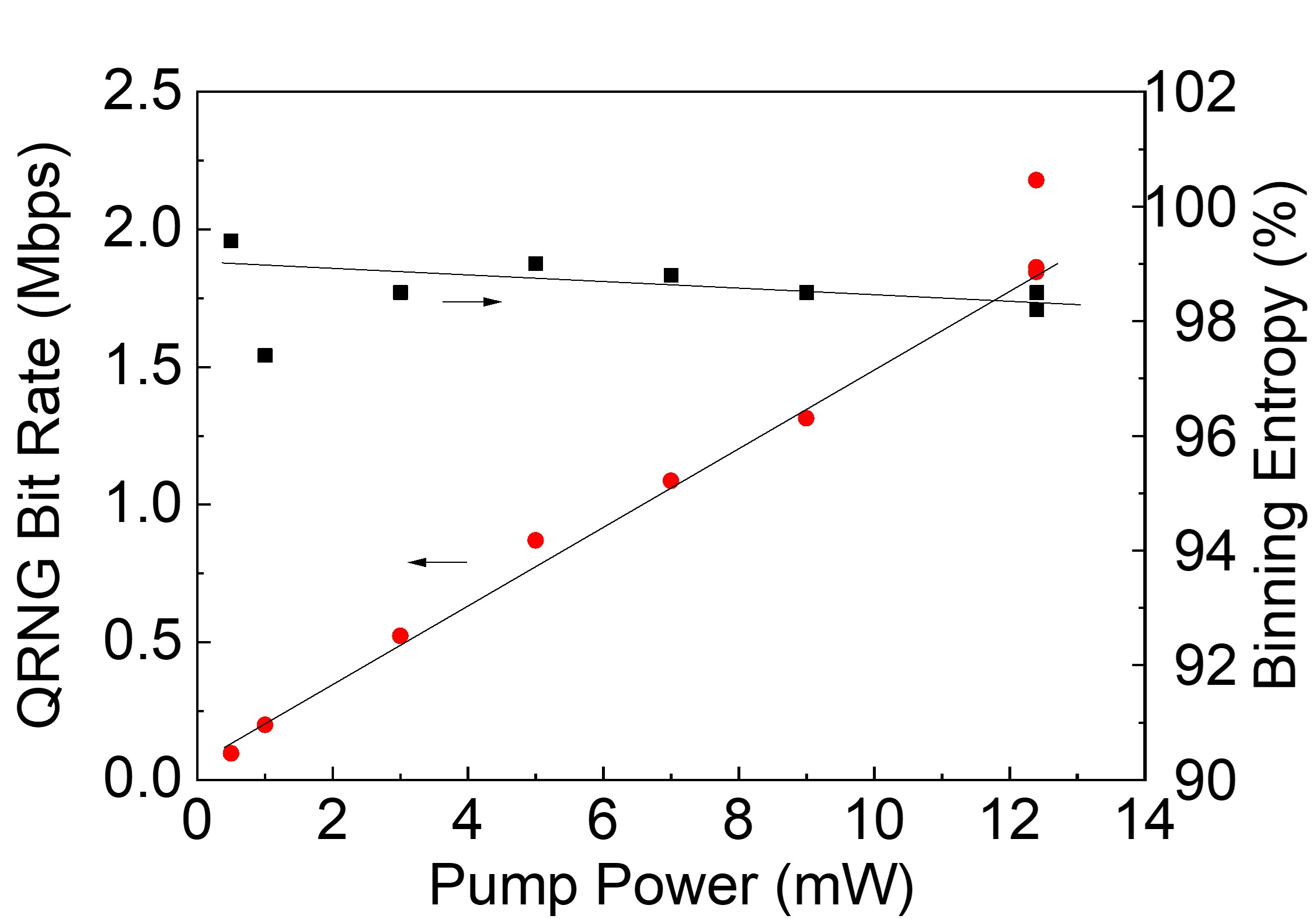}
    \caption{\textbf{Characterization of the QRNG system.} The variation of the minimum entropy, $H_{\infty}(X)$ (black dots), and post-processed QRNG bit rate (red dots), as a function of pump power and $g^{(2)}(0)$. The solid lines (red and black) are linear fit to the experimental data.}
    \label{Figure 3}
\end{figure}

\indent Knowing the performance metrics of the entangled photon source as a function of input power, we recorded the time stamps of the coincidence counts for pair-photons collected from the sections (U1 $\&$ D2) and (U2 $\&$ D1) and assigned binary values of 0 and 1, respectively, while simultaneously checking the Bell parameter of the source using the sections (C1 $\&$ C2) (see Fig.~\ref{Figure 1}). Subsequently, we recorded over 90 million raw bit sequences of 0’s and 1’s for each pump power. Due to the dependence of pump power on the generation rate of pair-photons, the recording time for 90 million bits decreased with increasing pump power, resulting in a recording time of 46.4 seconds at 12.4 mW of pump power for a coincidence window of 1 ns. Since the raw bit sequences contained all possible errors, we calculated the minimum entropy \cite{Xu:2012, Ma2013} as mentioned in our previous report \cite{Ayan:24} to estimate the maximum extractable unbiased true random bits. The results, presented in Fig.~\ref{Figure 3}, show that the min-entropy, $H_{\infty}(X)$ (black squares), remained nearly constant in the range of 99$\%$ to 97$\%$ as the pump power increased from 0.5 mW to 12.4 mW. For the increase of the coincidence window from 1 ns to 2 ns at a fixed pump power of 12.4 mW, the Bell parameter, $S$, decreased from 2.07 to 1.73 (see Fig.~\ref{Figure 2}(b)), indicating the loss of entanglement. However, it is interesting to note that there is no significant change in the min-entropy, $H_{\infty}(X)$, value even for S $<$ 2. 
Such observation demonstrates that the reduction in min-entropy is primarily due to classical noise in the experimental setup and is independent of the entanglement quality of the quantum source. To distill true random bits from the raw bit sequences, we employed a Toeplitz hashing randomness extractor \cite{krawczy}. The post-processed QRNG bit rate (red dots in Fig.~\ref{Figure 3}) exhibited a linear increase with pump power, rising from 0.09 Mbps at 0.5 mW to 1.8 Mbps at 12.4 mW, corresponding to a slope of 0.14 Mbps per mW. Additionally, at 12.4 mW pump power, increasing the coincidence window from 1 ns to 2 ns boosted the bit rate from 1.84 Mbps to 2.18 Mbps. These results confirm the realization of a high-bit-rate, device-independent QRNG system offering simultaneous Bell parameter measurement. Again, the higher value of min-entropy, $H_{\infty}(X)$, confirms the possibility for further enhancement of the bit rate of the QRNG system to a few tens of Mbps by simply increasing the pump power or by increasing the coincidence window. However, due to the unavailability of a higher power pump laser at 405 nm, we restricted the bit rate to 2.18 Mbps. Unlike our previous report \cite{Ayan:24}, the lower bit QRNG rate in the current experiment can be attributed to the larger ring diameter of the pair-photons to make easy division into six sections and lower effective power to the crystal.\\
%
\begin{figure}[H]
    \centering
    \includegraphics[width=\linewidth]{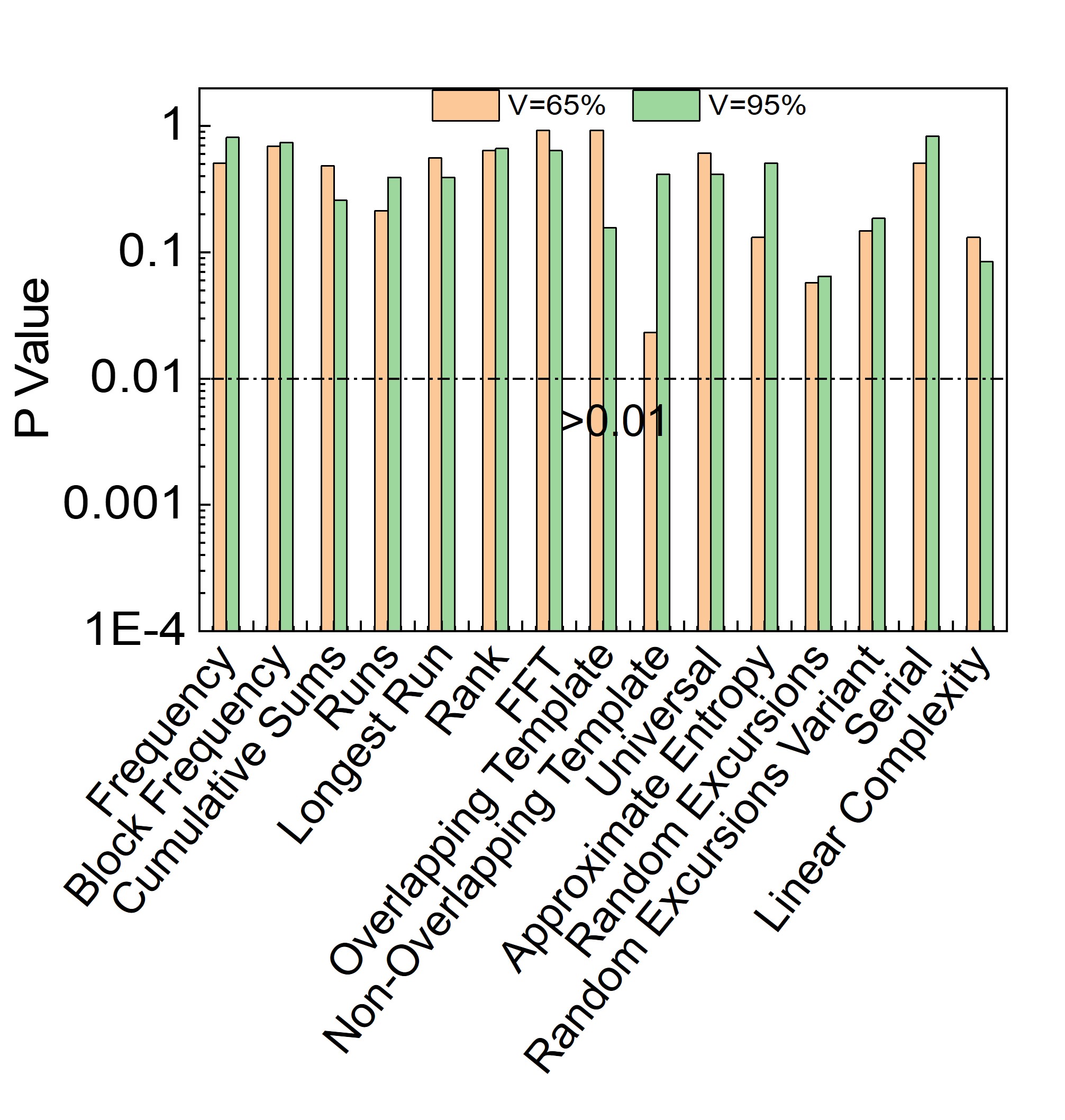}
    \caption{\textbf{Verification of randomness of the source through test suits.} Final P-values of 15 tests under NIST statistical suit for the Toeplitz post-processed bits recorded at 95\% and 61\% visibility, respectively.}
    \label{Figure 4}
\end{figure}
\indent To estimate the quality of randomness, we used the NIST 800-22 Statistical Test Suite \cite{rukhin} to the post-processed bit strings derived from the entangled photons with two different levels of entanglement visibility, V = 95 $\%$ and V = 65 $\%$. For each visibility setting, we divided the bit string into 80 sequences of binary bits, similar to our previous work \cite{Ayan:24}, each containing one million bits, thereby exceeding the minimum requirement of 55 sequences, as recommended by the NIST test suite. As the NIST statistical test suite follows two primary types of quantification to determine the randomness of the bit sequence, we calculated the uniformity of the P-values and the proportionality test to find the proportion of the input sequences passing (P-value is above the chosen significance level, $\alpha$, usually $\alpha$ = 0.01) a test. Using the Goodness-of-Fit distributional test and Kolmogorov-Smirnov tests to the post-processed bits, we calculated the final P-values of 15 different tests under the NIST test suite of the post-processed bits at visibilities, V = 95$\%$ (green) and V = 61$\%$ (orange), with the results shown in Fig. \ref{Figure 4}. As evident from the figure, the P-values of all the tests are $>$ 0.01, confirming the randomness of the experimental results at the different visibility levels. 
Although we have 80 sequences of bit strings, each consisting of 1 million bits for both entanglement visibilities of the quantum source, the NIST statistical test suite randomly selected $n$ = 46 sequences for the random excursions and random excursion variant tests. For these tests, using a significance level of $\alpha$ = 0.01 and $n$ = 46, the proportional fraction range was calculated to be (0.946, 1.034). For the remaining tests, with $\alpha$ = 0.01 and $n$ = 80, the proportional fraction range was determined to be (0.956, 1.023). We observed that the proportional fractions for all post-processed bit sequences fell within these specified ranges, confirming successful results across all tests and demonstrating the randomness of the generated bit sequences.

\indent Further, we evaluated the randomness of our post-processed bits using the TestU01 statistical test suite, specifically the Rabbit, Alphabit, and BlockAlphabit battery tests. While these tests typically require sequences of approximately 30 million bits, we employed longer sequences of 80 million bits for more rigorous analysis. Following the methodology outlined in the literature \cite{testu01, Neil}, we interpreted the P-values by considering values within the interval $\left[10^{-3}, 1-10^{-3}\right]$  as a success, and value outside this interval as a failure. We observed the P values for the post-processed bit string successfully passed all three tests, indicating the validation of our novel experimental scheme.\\
%

\begin{figure}[H]
    \centering
    \includegraphics[width=\linewidth]{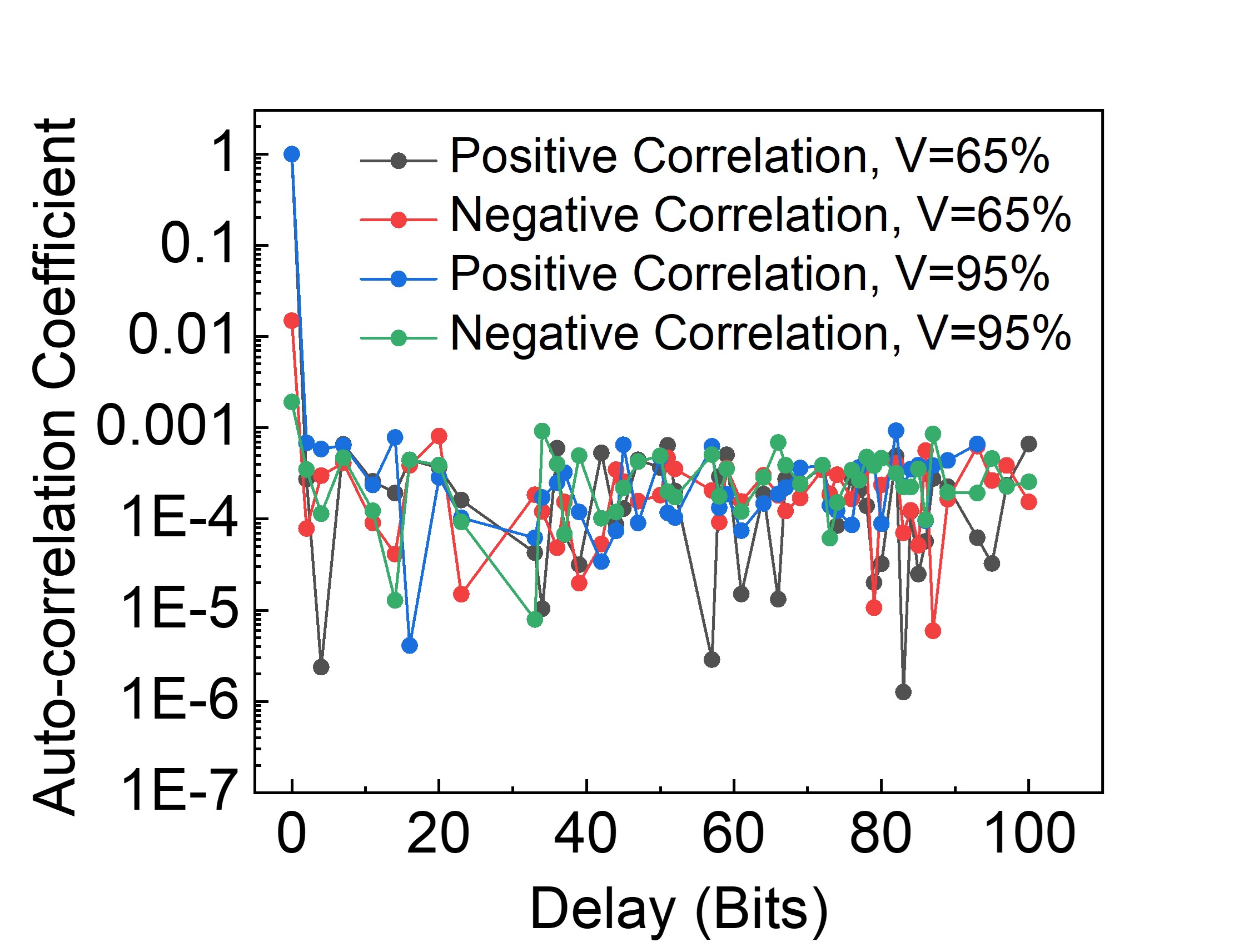}
    \caption{\textbf{Characterization of raw bits of the QRNG.} Variation of the magnitude of the positive and negative autocorrelation coefficients of the first 10 million bits of the raw bit sequence up to a delay of 100 bits. The raw bit sequences are recorded at two visibilities, 95$\%$ and 65$\%$.}.
    \label{Figure 5}
\end{figure}

\indent We further analyzed the randomness of the raw bits for two different entanglement visibilities, V = 95$\%$ (entangled state) and V = 61$\%$ (separable state), by calculating the autocorrelation coefficient. Using the first 10 million bits of the raw bit sequence, we computed the autocorrelation coefficient up to a delay of 100 bits, as shown in Fig.~\ref{Figure 5}. It is evident from the figure that even after a delay of 100 bits, the mean and standard deviation of the autocorrelation coefficient remain nearly unchanged. Given that a truly random sequence has an autocorrelation coefficient with a mean value close to zero, in our experiment, the raw bits for both entanglement visibilities exhibited a minimum autocorrelation coefficient in the order of $\sim$10$^{-6}$ and a standard deviation of \(9.8 \times 10^{-4}\). Such low values of minimum autocorrelation coefficient confirm that the raw bit string generated using the source of the entangled photons in the current novel experimental scheme represents a good random sequence even before post-processing, leading to a high min-entropy, \(H_{\infty}(X)\), as shown in Fig. \ref{Figure 3}, and enabling efficient extraction of high-quality random bits.

\begin{figure}[H]
    \centering
    \includegraphics[width=\linewidth]{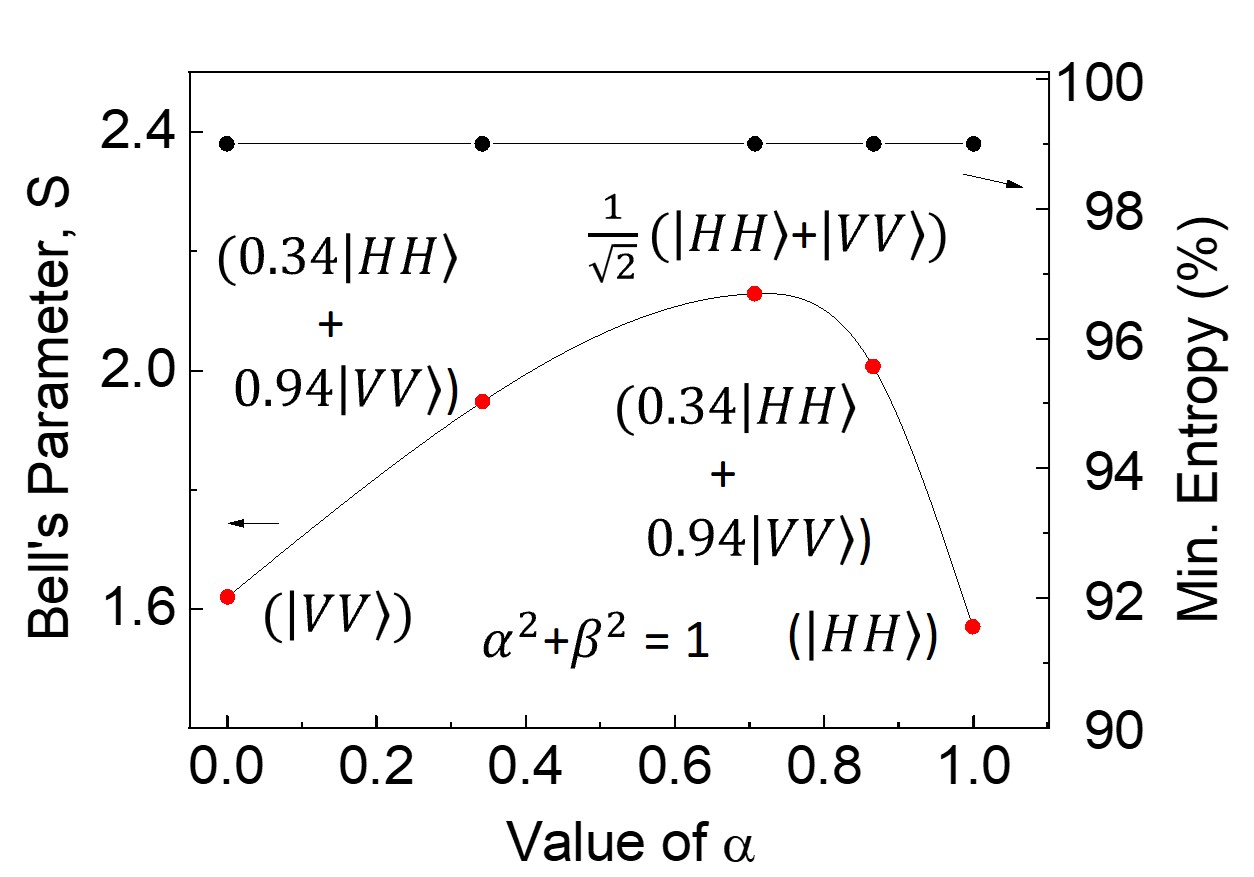}
    \caption{\textbf{Effect of output photon state on performance of QRNG.} 
    Variation of Bell's parameter and corresponding min-entropy of the raw bits of QRNG as a function of the weightage parameter, $|\alpha|$. The lines are guide to the eye.}
    \label{Figure 6}
\end{figure}

To validate the versatility of the current QRNG scheme, we further measured the min-entropy of the raw bits and the Bell's parameter, S, of the entangled photon state, while varying the coefficients, $|\alpha|$ and $|\beta|$. Keeping the pump power to the experiment constant, we varied the angle $\theta$ of the $\lambda/2$ plate from 0 to 45$\degree$ to change the power of the CW and CCW beams of the Sagnac loop and subsequently vary the output state of the pair-photons. As a result, the $|\alpha|$ (and corresponding $|\beta|$) value changes from 0 (1) to 1 (0), transitioning the output state of the pair-photons from $\ket{VV}$ to $\ket{HH}$. At $\theta$ = 22.5$\degree$, where $|\alpha| = |\beta| = 0.707$, the photons form a maximally entangled Bell state, $\ket{\phi^-} = (\ket{HH} - \ket{VV})/\sqrt{2}$. For other $\theta$ values, the state remains non-maximally entangled. The results, as presented in Fig. \ref{Figure 6}, show that the Bell's parameter, S (red dots), varies from 1.62 to 1.57, reaching a maximum of 2.13 for the maximally entangled state. Interestingly, the measured min-entropy ($\sim$99$\%$, black dots) of the raw bits collected for different states of the pair-photons remains consistent across different photon states, independent of the Bell's parameter, S. Again, for photon states, $\ket{VV}$ and $\ket{HH}$, and non-maximally entangled states, the value of Bell's parameter (S$<$2) might not certify the quantumness of the source used for QRNG. In such cases, one can use the $g^{(2)}(0)$, as used in our previous work \cite{Ayan:24}, as the suitable parameter to certify the quantumness of the source used for QRNG. This is because our novel QRNG scheme produces random bits based on the quantum-mechanical spatial and temporal correlation of the pair-photons.

To gain further insight into the current study, we have tabulated the bit rate and quantumness certification parameters of recently reported device-independent QRNG systems. As evident from Table \ref{table1}, our QRNG system, certified with a live Bell's parameter value, demonstrates a bit rate that is two orders of magnitude higher compared to previous reports.

\begin{table}[H]
\caption{\textbf{Comparison of the bit rate of QRNG based on coincidence detection.}}
\label{table1}
\centering
\small
\begin{tabular}{c c c c l} 
\hline
\hline
Reports & Certifying Parameter& Value & Bit rate  \\
  &  & & (bits/sec)\\
\hline \\
Ref.\cite{zhang2020}& S & - &1.71\\
Ref.\cite{liu2018} & S & - &$1.81 \cdot 10^2$\\
Ref.\cite{shen2018}& S & 2.016 & $2.40 \cdot 10^2$\\
Ref.\cite{Leone:2022} & S & 2.656 &$4.40 \cdot 10^3$\\
Ref.\cite{liu2021} & S & - &$1.35 \cdot 10^4$\\
Current study & S & 2.07 &$1.8 \cdot 10^6$\\
\hline
\hline
\end{tabular}
\end{table}

\section{Conclusions}
We have successfully demonstrated a novel beam-splitter-free, high-bit-rate DI-QRNG system with simultaneous quantumness certification through live Bell test data. The randomness is derived from direct access to the random occurrence of pair-photons at diametrically opposite points on the annular ring distribution of entangled photons, generated via a non-collinear degenerate SPDC process in a type-0 phase-matched PPKTP crystal within a polarization Sagnac interferometer. At a pump power of 12.4 mW and a 1 ns coincidence window, we achieved an entanglement visibility of $\sim$73$\%$ and a Bell parameter, $S$ = 2.07, and recorded high-quality quantum random bits of length of 90 million of raw bits having minimum entropy $>$97$\%$, which is further confirmed by an autocorrelation coefficient ($\sim 10^{-6}$). Using Toeplitz post-processing, we have verified the QRNG of a bit rate of 1.8 Mbps, passing all NIST and TestU01 test suites. Notably, increasing the coincidence window to 2 ns reduced visibility to 61$\%$ and the Bell parameter to $S$ = 1.73, but the produced raw bits has minimum entropy $>$95$\%$ and a post-processed QRNG bit rate of 2.18 Mbps. In the absence of the Bell parameter for the non-maximally entangled state or arbitrary output photon state, $g^{(2)}(0)$ can be the metric for the quantumness certification. Although we have divided the annular ring of the pair-photons into six sections, forming three high-brightness entangled photon sources from a single resource (laser, nonlinear crystal, and optical elements) for QRNG and live Bell's test, the generic experimental scheme can in principle be used to create more sections of the SPDC ring to form more number of entangled photon sources to increase the bit-rate of the Di-QRNG and for other advance quantum networks. The novel design simplifies the technical requirements of the DI-QRNGs and sets a new benchmark for the trustworthiness of the generated random numbers for real-world applications.

\section*{ACKNOWLEDGMENTS}
A. K. N., V. K., and G. K. S. acknowledge the support of the Department of Space, Govt. of India. A. K. N. acknowledges funding support for Chanakya - PhD fellowship from the National Mission on Interdisciplinary Cyber-Physical Systems of the Department of Science and Technology, Govt. of India through the I-HUB Quantum Technology Foundation.  G. K. S. acknowledges the support of the Department of Science and Technology, Govt. of India, through the Technology Development Program (Project DST/TDT/TDP-03/2022). M.E-Z acknowledges financial support from the Spanish Government through projects Ultrawave (EUR2022-134051), Alter (PID2023-150006NB-I00), Severo Ochoa Center of Excellence (CEX2019-000910-S); Generalitat de Catalunya (CERCA).

\section*{AUTHOR DECLARATIONS}
\subsection*{Conflict of Interest}
The authors have no conflicts to disclose.
\subsection*{Author Contributions}
A. K. N. developed the experimental setup and performed measurements. A. K. N. and V. K. participated in experiments, data analysis, numerical simulation, and data interpretation. M. E. Z. participated in the analysis and data interpretation. G. K. S. developed the ideas and led the project. All authors participated in the discussion and contributed to the manuscript writing.

\section*{DATA AVAILABILITY}
The data that support the findings of this study are available from the corresponding author upon reasonable request.

\bibliographystyle{IEEEtran}
\bibliography{Bib}


\end{multicols}
\end{document}